\documentclass[final]{elsart3p}

\usepackage{epsfig,pifont}

\usepackage{amssymb}
\usepackage{graphicx}
\usepackage{cite}

\begin{document}

\begin{frontmatter}



\title{Discriminant Analysis and Secondary-Beam Charge Recognition}


\author[1,9]{J.~{\L}ukasik\ead{j.lukasik@gsi.de}}, 
\author[1]{P.~Adrich},  
\author[1]{T.~Aumann}, 
\author[3]{C.O.~Bacri}, 
\author[4]{T.~Barczyk}, 
\author[5]{R.~Bassini}, 
\author[1]{S.~Bianchin}, 
\author[5]{C.~Boiano}, 
\author[1,6]{A.S.~Botvina}, 
\author[7]{A.~Boudard},  
\author[4]{J.~Brzychczyk},  
\author[8]{A.~Chbihi}, 
\author[9]{J.~Cibor}, 
\author[9]{B.~Czech}, 
\author[7]{J.-\'{E}.~Ducret}, 
\author[1]{H.~Emling}, 
\author[8]{J.~Frankland},  
\author[1]{M.~Hellstr\"{o}m}, 
\author[1]{D.~Henzlova}, 
\author[2]{G.~Imm\`{e}}, 
\author[5]{I.~Iori}, 
\author[1]{H.~Johansson},  
\author[1]{K.~Kezzar}, 
\author[3]{A.~Lafriakh}, 
\author[1]{A.~Le~F\`evre}, 
\author[7]{E.~Le~Gentil}, 
\author[1]{Y.~Leifels}, 
\author[1]{J.~L\"{u}hning}, 
\author[10]{W.G.~Lynch}, 
\author[1]{U.~Lynen},  
\author[4]{Z.~Majka}, 
\author[10]{M.~Mocko}, 
\author[1]{W.F.J.~M\"{u}ller}, 
\author[11]{A.~Mykulyak}, 
\author[2]{M.~De~Napoli}, 
\author[1]{H.~Orth}, 
\author[1]{A.N.~Otte},  
\author[1]{R.~Palit}, 
\author[9]{P.~Paw{\l}owski}, 
\author[5]{A.~Pullia}, 
\author[2]{G.~Raciti}, 
\author[2]{E.~Rapisarda}, 
\author[1]{H.~Sann\thanksref{dec}}, 
\author[1]{C.~Schwarz}, 
\author[1]{C.~Sfienti}, 
\author[1]{H.~Simon}, 
\author[1]{K.~S\"{u}mmerer}, 
\author[1]{W.~Trautmann}, 
\author[10]{M.B.~Tsang}, 
\author[10]{G.~Verde}, 
\author[7]{C.~Volant}, 
\author[10]{M.~Wallace}, 
\author[1]{H.~Weick}, 
\author[1]{J.~Wiechula}, 
\author[4]{A.~Wieloch}  \and
\author[11]{B.~Zwiegli\'{n}ski}
\thanks[dec]{deceased}


\address[1]{Gesellschaft f\"ur Schwerionenforschung mbH, D-64291 Darmstadt, Germany}
\address[2]{Dipartimento di Fisica e Astronomia dell'Universit\`a and INFN-LNS
and Sez. CT, I-95123 Catania, Italy}
\address[3]{Institut de Physique Nucl\'eaire, IN2P3-CNRS et Universit\'e, F-91406 Orsay, France}
\address[4]{M. Smoluchowski Institute of Physics, Jagiellonian University, Pl-30059 Krak\'ow, Poland}
\address[5]{Istituto di Scienze Fisiche, Universit\`a degli Studi and INFN, I-20133 Milano, Italy}
\address[6]{Institute of Nuclear Research, Russian Accademy of Science, Ru-117312 Moscow, Russia}
\address[7]{DAPNIA/SPhN, CEA/Saclay, F-91191 Gif-sur-Yvette, France}
\address[8]{GANIL, CEA et IN2P3-CNRS, F-14076 Caen, France}
\address[9]{IFJ-PAN, Pl-31342 Krak\'ow, Poland}
\address[10]{Department of Physics and Astronomy and NSCL, MSU, East Lansing, MI 48824, USA}
\address[11]{A. So{\l}tan Institute for Nuclear Studies, Pl-00681 Warsaw, Poland}

\begin{abstract} 

The discriminant-analysis method has been applied to optimize the exotic-beam
charge recognition in a projectile fragmentation experiment. The experiment was
carried out at the GSI using the fragment separator (FRS) to produce and select
the relativistic secondary beams, and the ALADIN setup to measure their
fragmentation products following collisions with Sn target nuclei. The beams of
neutron poor isotopes around $^{124}$La and $^{107}$Sn were selected to study
the isospin dependence of the limiting temperature of heavy nuclei by comparing
with results for stable $^{124}$Sn projectiles. A dedicated detector to measure
the projectile charge upstream of the reaction target was not used, and
alternative methods had to be developed. The presented method, based on the
multivariate discriminant analysis, allowed to increase the efficacy of charge
recognition up to about 90\%, which was about 20\% more than achieved with the
simple scalar methods.

\end{abstract}

\end{frontmatter}

\section{Introduction}

The ALADIN experiment S254, conducted in 2003 at the SIS heavy-ion synchrotron 
at GSI Darmstadt, was designed to study isotopic effects in projectile 
fragmentation at relativistic energies. Besides stable $^{124}$Sn beams,
neutron-poor secondary Sn and La beams were used in order to extend the range
of isotopic compositions beyond that available with stable beams alone. The
secondary beams were produced  at the fragment separator FRS \cite{frs92} by
the fragmentation of primary $^{142}$Nd  projectiles with energies of about 900
MeV per nucleon in a thick beryllium  target. The FRS was set to select
$^{124}$La and, in a second part of the experiment, also $^{107}$Sn secondary 
projectiles of 600 MeV per nucleon which were then delivered to the experiment.

\begin{figure}[htb!]
\centering
\includegraphics[width=0.95\columnwidth]{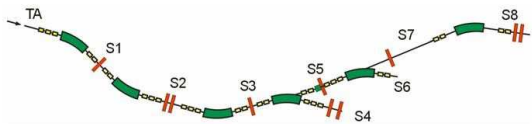}

\caption{Schematic FRS beam line setup. The primary beam from SIS enters the
production target area (TA) from the left. S1-S8 - focal planes. S2 -
dispersive focus, S8 - secondary beam exit to Cave B and C (after
http://www-w2k.gsi.de/frs-setup).}

\label{lff}
\end{figure}

\begin{figure}[htb!]
\centering
\includegraphics[width=0.95\columnwidth]{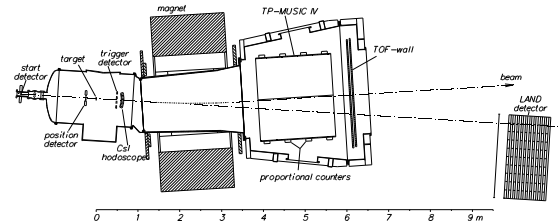}

\caption{S254 setup in Cave B. The secondary beams from the FRS enter the
experimental area from the left (see Ref. \protect{\cite{sfienti05}}).}

\label{lfa}
\end{figure}

In order to reach the necessary beam intensity of about 1000 particles/s  with
the smallest possible mass-to-charge ratio $A/Z$, it was found necessary  to
reduce the thickness of the degrader at the dispersive focus (S2) of the FRS
(see Fig. \ref{lff}) that is normally used for the selection of isotopically
pure secondary beams. As a result, the beams arriving at the reaction target
contained a distribution of neighbouring isotopes together with the nominal
$^{124}$La or $^{107}$Sn. The $A/Z$ of each projectile is known from the
measured position at the dispersive focus (S2) and from its flight time over
the more than 80 m beam line between the plastic scintillator at S8 (FRS exit)
and the start detector at the entrance of the ALADIN setup (Figs. \ref{lff} and
\ref{lfa}). The atomic number $Z$, however, could only be measured for
non-interacting projectiles with the MUSIC IV tracking detector placed
downstream of the ALADIN dipole magnet.


In order to minimize the upstream material, only two thin plastic detectors 
were mounted near the reaction target (Fig. \ref{lfa}): the start detector, for
measuring the arrival time of each projectile, and the position detector, for
measuring the projectile position in the plane perpendicular to the beam
direction. No detector with the resolution necessary to resolve  individual
$Z$'s in the $Z\approx 50$ region was placed there. It will be shown in the
following that, with a discriminant method, the atomic number $Z$ of
interacting projectiles can, nevertheless, be reconstructed with high
resolution by combining the amplitude and time information available from all
the upstream detectors (S2, S8, start and position).

\section{Discriminant analysis}

Discriminant analysis (DA) is a statistical method for predicting a class
membership for a set of vectors of observations based on a set of observations
for which the classes are known (a training set). For the training set, DA
searches for a transform -- a projection onto a lower-dimensional vector space
-- such that the ratio of the between-class distance to the within-class
distance is maximized, thus assuring maximum discrimination. The optimal
reference frame onto which the projection is made, is found by solving the
generalized eigenvalue problem for the scatter matrices (see e.g. \cite{jain88,
ye04, desesquelles95} and their references) defined for the training set. For a
training set composed of $K$ classes, each $n_{k}$ in number, the total, 
${\cal S}_{T}$, and the within-class, ${\cal S}_{W}$, scatter matrices can be
defined using the notion of covariance matrix:

\[{\cal S}_{T} = N \mbox{\boldmath$C$}_{T},\]

\[{\cal S}_{W} = \sum_{k=1}^{K} n_{k} \mbox{\boldmath$C$}_{W}^{k},\]

\noindent and the between-class scatter matrix, ${\cal S}_{B}$, can be simply
expressed as a difference\footnote{See e.g. \cite{jain88,ye04} for a strict
definition}:

\[{\cal S}_{B} = {\cal S}_{T}-{\cal S}_{W}.\]

Here, $N = \sum_{k=1}^{K} n_{k}$, is the total number of vectors of
observations in the training set and $\mbox{\boldmath$C$}_{T}$ and
$\mbox{\boldmath$C$}_{W}^{k}$ are the covariance matrices, built of the vectors
of observations, for the total set and for each individual class, $k$,
respectively.

The optimal reference frame is found as a sub-set of eigenvectors $\vec v$
solving the generalized eigenvalue problem:

\begin{equation}
{\cal S}_{B} \vec v = \lambda {\cal S}_{T} \vec v.
\label{eigprob}
\end{equation}

An important property of DA is its dimensionality reduction. Since the rank of
${\cal S}_{B}$ cannot be greater than $(K-1)$ and the rank of a product cannot
exceed the ranks of the components, the optimal frame, in which the classes are
best separated, is spanned by at most $(K-1)$ eigenvectors. This means, for
example, that for a 3-class training set considered here (see next section) and
189 dimensional vectors of observations, the problem gets reduced after DA to 2
dimensions, such that the structure of the original high-dimensional space is
preserved in the low-dimensional space. The problem can then easily be handled
with {\em e.g.} 2 dimensional histograms. It should be noted that, to avoid the
singularity problem, the initial dimensionality should not exceed the number of
vectors of observations \cite{ye04}.

DA has a broad range of applications: from its pioneering application in plant
taxonomy \cite{fisher36}, through applications {\em e.g.} in image
recognition,  marketing, diagnosis in medicine, to those in nuclear physics
\cite{desesquelles95, desesquelles95a}. We found it also very well suited for
recognizing the charge of secondary beams from the GSI fragment separator used
as projectiles in the S254 experiment \cite{sfienti05}. In order to minimize
the background from the reactions up-stream, the ionization chambers, normally
used in secondary-beam experiments to determine the Z of the beam, were not
used in S254. The charge of the beam entering Cave B could only be measured
precisely for the projectiles which did not interact with the target, using the
MUSIC IV tracking chamber positioned down-stream of the target (see Fig.
\ref{lfa}). In order not to overload the data stream with these
non-interacting-beam events, only a selected fraction of them has been allowed
to trigger (after scaling down) and these events were used as the training set.

\section{Results and discussion}

From realistic simulations of the beam transport, it was found that the most
promising observable to determine the up-stream charge is the position of the
beam at the target. During the experiment this was measured with the
position-sensitive plastic scintillator detector required for the tracking of
projectile fragments and positioned just in front of the target (see Fig.
\ref{lfa}). Optionally, some other simple observables could be considered as
beam charge sensitive as, {\em e.g.}, the energy loss in the plastic detectors
up-stream (S2, S8, start or position, see Fig. \ref{lfa}) or the $A/Z$
variation with the $Z$ of the beam.

\begin{figure}[htb!]
\centering
\includegraphics[width=0.8\columnwidth]{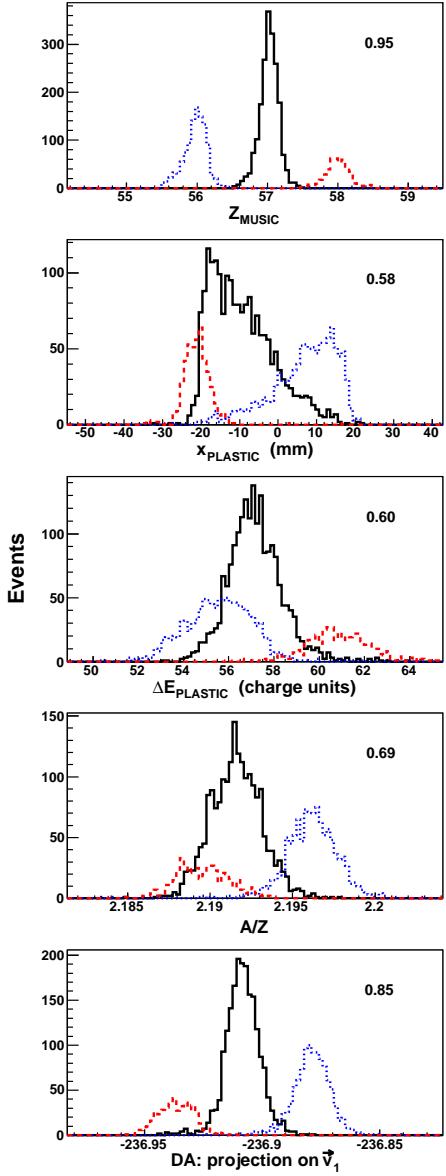}

\caption{(Color online) Charge distributions for $Z=57\pm 1$ beams of the
training set obtained from the (from top to bottom) energy loss in the MUSIC
chamber, horizontal position of the beam at the target, energy loss $\Delta
E_{\rm PLASTIC}$ in the position detector, high resolution $A/Z$ and the DA
with 189 dimensional input vectors of observations projected on the 
eigenvector corresponding to the larger eigenvalue (bottom, see Fig.
\ref{lf0}). The numbers (top-right) specify the discrimination power
\cite{desesquelles95a} of the observable. The spectra are plotted in the range
of $\pm 4$ standard deviations of the total. The solid ($Z=57$), dotted
($Z=56$) and dashed ($Z=58$) histograms represent three classes of the beam
charge labeled according to the energy loss in the MUSIC chamber.}

\label{lf1}
\end{figure}

All these beam charge correlated scalar observables, together with the
``reference'' energy loss in the MUSIC chamber and the final
multidimensional-DA projection, have been tested for their charge
discrimination power. Figure \ref{lf1} presents distributions of these
observables for the training set of secondary beams of $Z=57\pm 1$, for a
selected $A/Z$ interval $[2.183,2.203]$ containing the $^{125}$La isotope, and
plotted in the range of $\pm 4$ standard deviations from the mean of the total
spectrum. The three histograms in each panel represent three classes of the
beam charge ($Z=56$ - dotted histogram, 57 - solid and 58 - dashed) and are
labeled according to the charge identification derived from the energy loss in
the MUSIC chamber.

The top panel of Fig. \ref{lf1} shows a ``reference'' charge distribution of
secondary beams as measured by the MUSIC chamber for non-interacting
projectiles from the same run. This observable served to  label the classes of
the training set.  It is unavailable for projectiles that underwent
fragmentation in the target but, from the experimental point of view, of
interest for analyses  requiring a precise knowledge of the atomic number $Z$
besides the known $A/Z$.

The three middle panels present the corresponding ``charge distributions'' 
obtained by measuring the horizontal position of the beam at the target (corresponding to the dispersive coordinate of the FRS), the energy loss
in the position detector, and the high resolution $A/Z$.

\begin{figure}[htb!]
\centering
\includegraphics[width=0.8\columnwidth]{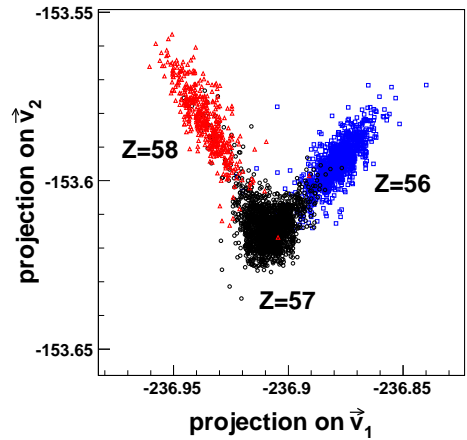}

\caption{(Color online) Projections of the original 189-dimensional vectors of
observations on the two main eigenvectors ($\vec v_{1}, \vec v_{2}$) obtained
with the DA method for the 3-class training set. The three clouds of symbols
represent three classes of charge of the example training set: $Z=57$ (black
circles), $Z=56$ (blue squares) and $Z=58$ (red triangles).}

\label{lf0}
\end{figure}

The bottom panel of the figure shows the same three classes of charge plotted
as projections on the eigenvector corresponding to the largest eigenvalue,
obtained using the DA technique with 189 dimensional input vectors of
observations (see Fig. \ref{lf0} for a corresponding 2-dimensional original
projection). The components of these input vectors of observations (features)
contain the information from all the up-stream detectors, {\em i.e.} the times
and amplitudes from the S2 and S8 FRS scintillators \cite{frs92}, the start
detector and from the position detector in front of the target. All together 18
raw signals which were combined in addition into pair products to maximize the
DA efficiency (see {\em e.g.} \cite{desesquelles95a}), finally resulting in 189
features.

The figure demonstrates that the high charge resolution obtained with the MUSIC
chamber is difficult to be achieved or surpassed with other simple observables. However,
combining all the possible pieces of information in an optimal way using the DA
method allows to improve the resolution and to come much closer to the reference.

One way to quantify the resolution of an observable is to determine its 
discrimination power. It is defined as \cite{desesquelles95a}:

\[ \lambda=\frac{{\cal S}_{B}}{{\cal S}_{T}} \]

\noindent for 1-dimensional vectors of observations (scalar observables)
and as an appropriate eigenvalue of (\ref{eigprob}) for the multidimensional
case. $\lambda \in [0,1]$ and is equal to zero when the classes have the same
mean values and equal to one when, for all classes, all the vectors of
observations of a given class are equal but distinct from those of the other
classes. Discrimination powers for the tested observables have been specified
in the top-right corner of each panel of Fig. \ref{lf1}.

The discrimination among classes becomes even better since in actual
discrimination procedures we consider at least three classes at a time and thus
deal with at least two-dimensional projections. Fig. \ref{lf0} shows the
original 2-dimensional scatter plot for the three classes of charge considered
here which was the base for the one-dimensional projection presented in the
bottom panel of Fig. \ref{lf1}. The distribution results from the projection of
the original 189-dimensional vectors of observations on the two main
eigenvectors obtained with the DA technique.

Having a way to label the classes of the training set and the DA tool to define
a low-dimensional space in which the classes are best separated, one can now
project the vectors of observations of unknown class membership onto the
optimal reference frame and, using the distributions of the training set,
assign to them the most probable (or, if possible, average) labels. 

In our application the label (charge) assignment has been done in two ways.
Since the beam charge range is quite narrow, we restricted the individual
charge identification to  5 classes of charges with $Z \in [Z_{nom} - 2,
Z_{nom} + 2]$ (with nominal charge $Z_{nom} =$ 57 (50) for the nominal
$^{124}$La ($^{107}$Sn) beam) plus two classes with $Z < Z_{nom} - 2$ and $Z >
Z_{nom} + 2$, thus altogether 7 classes. In the first approach this definition
of classes has been subsequently simplified by splitting the 7 classes into 5
groups, each group containing only 3 classes: $Z_{0}$ with $Z \in [Z_{nom} - 2,
Z_{nom} + 2]$, $Z_{<}$ with $Z < Z_{0}$ and $Z_{>}$ with $Z > Z_{0}$. This
allowed, thanks to the dimensionality reduction of DA, to base the charge
assignment on the usage of 2-dimensional histograms, such as the one in Fig.
\ref{lf0}, defined using the vectors of the training set.

The other approach was based on the original 7-class definition which lead to a
6-dimensional optimal space after DA. The 6-dimensional distributions of
projections of the training vectors were then parametrized assuming their
Gaussian shapes and these parametrizations were used for the charge assignment.
The first approach, based on the 2-dimensional reference histograms, was found
to give about 2\% better efficacy of charge recognition as compared to the
method based on the Gaussian parametrization. 

\begin{figure}[htb!] \centering
\includegraphics[width=0.8\columnwidth]{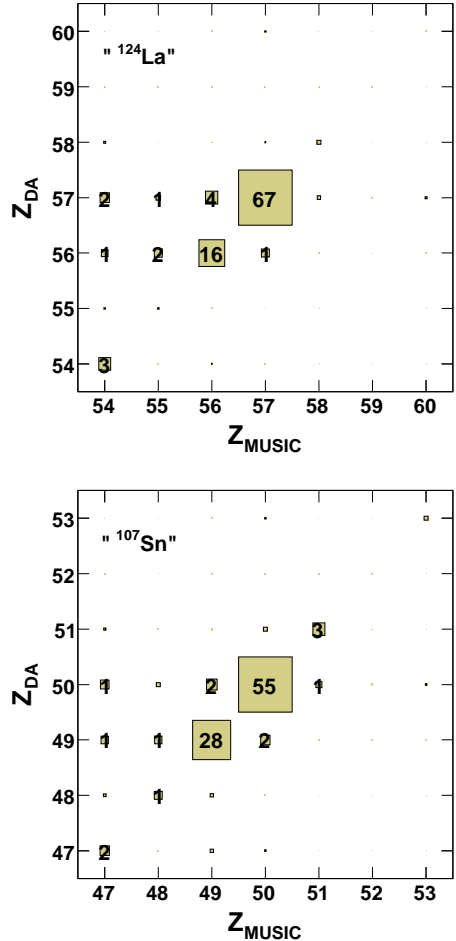}

\caption{Contributions of charges reconstructed using the DA-based method
$(Z_{DA})$ and the charges measured by the MUSIC chamber $(Z_{MUSIC})$. The
numbers in boxes represent the contributions in \% of the total yield. The
first (last) bin contains also the contributions of all the smaller (greater)
charges.}

\label{lf_eff}
\end{figure}

Fig. \ref{lf_eff} shows the proportion of charges reconstructed using the DA
method, $Z_{DA}$, as a function of the reference charge measured by the MUSIC
chamber, $Z_{MUSIC}$. The numbers in boxes represent the proportions, in \% of
the total yield, calculated for the total statistics collected. They are
indicated if the contribution exceeds 1\%. The sum of the diagonal elements
provides the measure of the overall efficacy of the method, {\em i.e.} the
proportion of correctly recognized charges. It amounts to about 86\% in case of
the nominal "$^{124}$La" beam and to about 89\% for the "$^{107}$Sn" beam. The
first and the last bins contain additionally the contributions of all the
smaller and greater charges, respectively. In particular, the figure shows that
the light impurities formed in reactions on the up-stream material (the first
column) amount to about 4-6\% of the incoming projectiles. About 50\% of them
can be properly recognized and eventually rejected during the analysis. The
methods based on the simple scalar observables have been found to be about
20\% less efficacious than the one based on DA.

The ratio of the mass to atomic numbers, $A/Z$, of the projectile has been
measured with high precision from the time of flight and position at the
dispersive focal plane S2 of the FRS. The calibration has been done using
the basic equation for identification: 


\[
B\rho = \frac{A}{Z} \frac{u \beta}{e \sqrt{1-\beta^2}} 
\]

\noindent where, $B\rho$ is the projectile rigidity and $A$, $Z$ and $\beta$
its mass and charge numbers and velocity, respectively; $u$ and $e$ are the
atomic mass and charge units.

\begin{figure}[htb!]
\centering
\includegraphics[width=0.75\columnwidth]{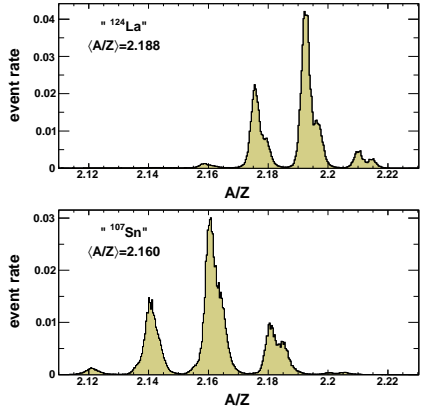}

\caption{Distributions of $A/Z$ for "$^{124}$La" (top) and "$^{107}$Sn"
(bottom) nominal beams. The numbers in the top left corners specify the average
$A/Z$ of the secondary beams for the whole experiment.}

\label{lf2}
\end{figure}

Figure \ref{lf2} presents the measured distributions of $A/Z$ for both
secondary beams used in the experiment. The high precision measurement allows
to visualize the ``fine'' structure of the $A/Z$ groups due to its $Z$
dependence. The average $\langle A/Z \rangle$ for the two secondary beams are:
2.188 for the nominal "$^{124}$La" beam and 2.160 for the "$^{107}$Sn" beam.
The average $\langle Z \rangle$ for the two beams are 56.8 and 49.7,
respectively. The averages take into account variations due to different FRS
settings for different groups of runs and are taken over the whole experiment.

\begin{figure}[htb!]
\centering
\includegraphics[width=0.8\columnwidth]{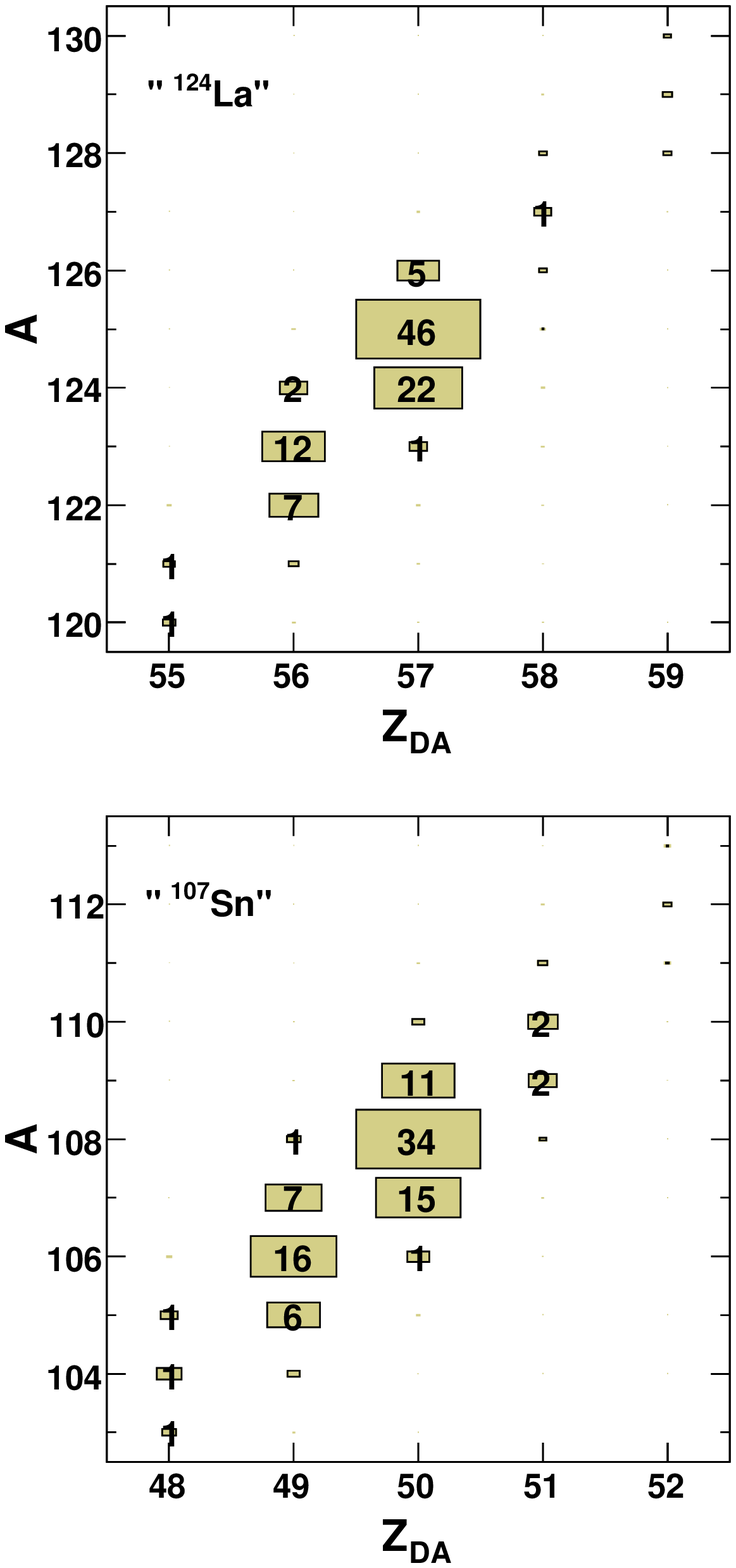}

\caption{Mass versus atomic number of the two secondary beams used in the S254
experiment, as obtained from the $Z$ reconstruction using the DA method and the
high precision $A/Z$ measurement. The numbers in the boxes represent the
contribution of a given isotope (in \%) to the secondary beam.}

\label{lf3}
\end{figure}

Having determined the $A/Z$ and $Z$, the isotopically pure beams can be
selected also for events with reacting projectiles. Figure \ref{lf3} shows the
resulting composition of the two secondary beams used in the S254 experiment.

The numbers in the histograms specify the contribution of a given isotope (in
\%) and are indicated only if the contribution is greater than 1\%. Gating on a
given atomic and mass number allows to select a 90\% pure beam and to also keep
away from the light impurities from reactions on the up-stream material.

Summarizing, we have demonstrated a high efficacy of the secondary-beam charge
recognition method based on the discriminant analysis technique in an
experiment without a dedicated Z-detector. The method has improved the quality
of the S254 experimental data by allowing up to 90\% pure isotopic selection of
the incoming projectiles.


The authors would like to thank P. D{\'e}sesquelles for stimulating
introduction to DA and valuable comments. This work has been supported by the
Polish Ministry of Science and Higher Education under Contracts No. 1 P03B 105
28 (2005 - 2006) and N202 160 32/4308 (2007-2009).

\vspace{-3mm}



\end{document}